# INFORMATION SECURITY PLAN FOR FLIGHT SIMULATOR APPLICATIONS

Jason Slaughter[1] and Syed (Shawon) M. Rahman, PhD[2]

[1]Information Assurance and Security, Capella University, Minneapolis, USA
jslaughter1@capellauniversity.edu
[2]Assistant Professor, University of Hawaii-Hilo, Hilo, USA
and Adjunct Faculty, Capella University, Minneapolis, USA
SRahman@Hawaii.edu

## ABSTRACT

*The Department of Defense has a need for an identity management system that uses two factor authentications to ensure that only the correct individuals get access to their top secret flight simulator program. Currently the Department of Defense does not have a web interface sign in system. We will be creating a system that will allow them to access their programs, back office and administrator functions remotely. A security plan outlining our security architecture will be delivered prior to the final code roll out. The plan will include responses to encryption used and the security architecture applied in the final documentation. The code will be delivered in phases to work out any issues that may occur during the implementation.*

## Keywords

*Information Security, Public Key Encryption, Certificate Authentication, Token Authentication, Identity Management, Internet Security*

## 1. INTRODUCTION

The Department of Defense has a need for a secure sign in system for their Flight Simulator program. The program will allow the war fighter to access the latest flight combat strategies from anywhere in the world. The Department of Defense has contracted XYZ Security to implement a mixed COTS (Commercial off the shelf) and custom code solution. The customer wants a web facing solution that will allow their internal users, contractors and any additional personnel that meet their requirements to gain access to the back end application. The customer considers this program to be a crucial asset [30] that should be protected from unauthorized individuals. The DoD has taken an initial look at creating a cloud computing [31] solution. However, they feel that the risks are still high.

They would like us to deliver a solution that would allow them to keep all of the hardware in house. The DoD has asked us while implementing our solution to keep in mind suggestions on how they can improve their security risk management [32]. Specifically they would like to create a stronger relationship between their physical security policies and their information security policies. As an initial deliverable the customer would like to see a security plan that meets all of the high level requirements they have requested as well as meeting industry best practice for a secure enterprise web application.

They would like us to take into account current wiretapping legislation [33] and ensure that none of our solutions create a situation where it may be perceived that our sign in solution is creating a wire tapping situation in appearance or in fact. We are currently working with senior architects to verify our solution will not create this issue.

The solution must meet the following requirements.
- FISMA Compliance [3]
- NIST Level 4 Compliance [4]
- Two-Factor Authentication [20]
  - Using Tokens
  - Using CAC Cards
- Identity Management
- Non-Repudiation
- PKI Certificate Authentication [4]
- Deliverable: Analysis of Encryption for Implementation.
- Deliverable: Compare Cryptographic Methods. Define Strengths and Weaknesses. Defend Final Decision.
- Deliverable: Analysis of Various Cipher Attacks. Outline Mitigation Plan.
- Deliverable: Evaluate SDLC. Show how the change control system will ensure the strength of the final deliverable.
- Analysis of the database. Outline the best security practices used around the implementation.

## 2. CASE STUDY: PROMETHEUS FLIGHT SIMULATOR

The Prometheus Flight Simulator project was the pilot project for this endeavour. The customer had a need to create a robust system that would meet their security needs and at the same time create a reliable process their user base would be willing to accept.

### A. Organizational Structure

The government organization has a "RED" team compromised of five individuals who will act as the subject matter experts moving forward. The team will be responsible for coordination with the end user group. The end user group will at first be composed of one hundred IT professionals across various IT disciplines. These individuals will act as the acceptance team for the solution. This team currently uses a role based system to outline ACL's [2] for each team member. The access control list will have to be updated and referenced throughout the software development lifecycle. This list should be the central authority on user access.

Going forward the "RED" team and the end users will be referred to as the customer. We will be using an agile approach to make sure the customer stays in the loop throughout the entire software development lifecycle. Our external organization is comprised of ten individuals with an average of twenty years experience across the board. Our team will be responsible for Architectural Design, Implementation, Unit Testing, Training and Environmental Setup. The customer will be responsible for Change Management, Acceptance Testing, Training and Requirements Finalization.

System administrators will be identified by the "RED" team. Each of these administrators will work with the development team during environment setup. The administrators will gain hands on experience with the new system and make them the owner of that system moving forward. At a production level the customer expects to be able to on-board people as needed. An on-boarding system will have to be implemented to the administration sever for the customer. This is currently a phase two requirement. In phase one the user base will only be comprised of the currently defined customer make up of one hundred IT professionals and the "RED" team. More personnel may be on-boarded at their discretion. However, it is out of scope in phase one for our organization to create this ability. The phase three final deliverable will have the capability to on-board up to one hundred thousand end users. This should be kept in mind during the development effort.

## B. Security Architecture

Currently the Department of Defense relies on decentralized security architecture to allow for high availability of their systems with a low impact when one the decentralized systems go down. An example of one of these systems is the current Distributed Meta Data Repository [1]. Due to the size of the organization the systems have multiple administrators and many administrators may be needed to troubleshoot problems based on where they originate and the systems affected.

Moving forward our team has chosen to follow the current accepted security architecture. We will design a decentralized system that will use several subsystems to comprise the entire solution.

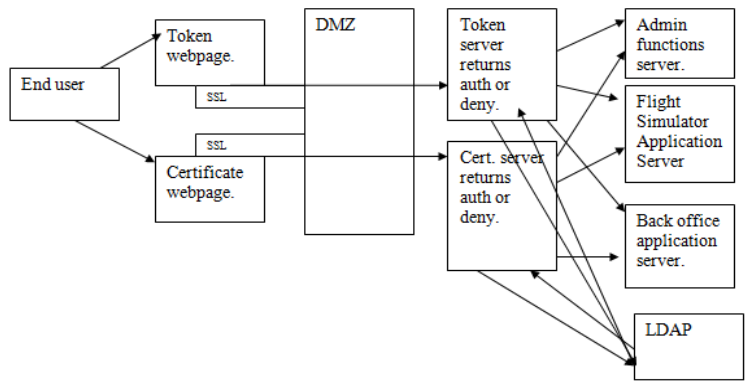

Fig. 1 – Security Architecture Design Diagram: Prometheus Flight Simulator

The diagram above outlines the user sign in [Fig. 1] at a high level. While the diagram does not go into low level details, it serves as a reference regarding the design of the architecture and the layers the user will go through from their client to the back end server. The solution will be tied together by internal networks. The internal networks will be accessed from Internet by a single sign on through Public Key Encryption or Token Authentication.

## C. Identified Systems

The flight simulator application has several identified systems that require an assessment and evaluation. The systems will act as the primary layers of security between the protected data and the outside world. While the application itself is considered a protected resource, it also acts a layer of protection between the user and the database.

- PKI Server
- Token Authentication Server
- DMZ and Network
- Flight Simulator Application and Server
- Administrator Application and Server
- Back Office Application and Server

### D. **Public Key infrastructure**

The requirements outline a need for the Prometheus Flight Simulator to support a Public Key Infrastructure sign in process to gain access to the system. The customer currently has several common access cards in use for other systems that we may be able to leverage moving forward. Because of the nature of the project we cannot leverage the existing certificate authority. This document will outline the plan for creating a certificate authority for the Prometheus Flight Simulator project.

### E. **Public Key infrastructure: High level requirements**

A public key infrastructure will be one of the two main authentication methods used when the user authenticates to the application. The security plan design must meet the following high level requirements.
- Establish Trust
- Analysis of why trust is important in disguising and protecting data.

## II. **E.1. Public Key infrastructure: trust**

A PKI (Public Key Infrastructure) [4] sign in happens when a user clicks a link to a site or resource and the server asks the client for a certificate. The user picks the certificate loaded in their browser; in this case the certificate will come from the CAC (Common Access Card) [3]. The certificate will be compared against the list received from the certificate authority to determine if it is a valid certificate. If the certificate is valid and has not been revoked the user will be allowed into the system. The certificate authority is sole creator of trust in the certificate. An assessment of whether to trust the certificate authority must always be conducted.

## III. **E.2. Disguising and protecting data**

Public Key Infrastructure uses asymmetric cryptography to relay data back and forth between two parties who may not know each other [Fig. 2], but must authenticate with each other to exchange data securely. Using the PKI system the public keys are used to encrypt a message and the private keys are used to decrypt messages. The certificate authority verifies that the public keys actually belong to who they says they do. This creates a system where confidentiality, integrity and accountability exist for both users. In this system it is possible to also verify that tampering has not occurred by asking the sender of the message to also send a hash of the plain text message. The receiver will then rehash the sent message to verify the hashes match [5].

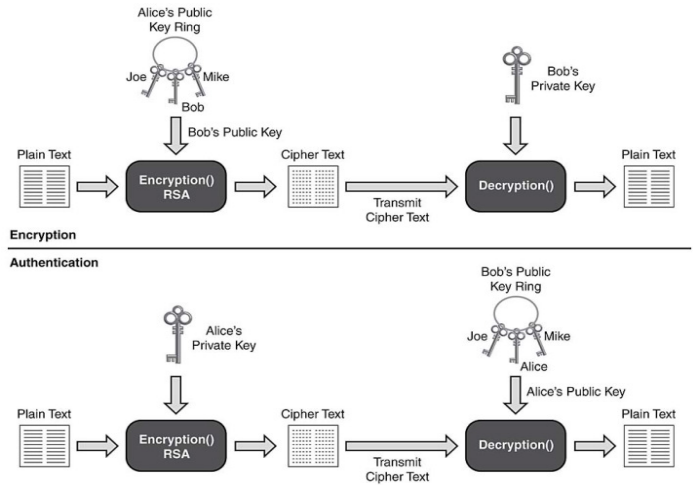

Fig. 2 – Asymmetric Encryption [10]

## IV. E.3. PKI Authentication

According to the requirements the Prometheus Flight Simulator shall allow a user to sign in using a Public Key Infrastructure sign in system. We have used the following implementations to meet this requirement.

Public Key Infrastructure uses asymmetric cryptography to relay data between two parties who may not know each other, but must authenticate with each other to exchange data securely. Using the PKI system the public keys are used to encrypt a message and the private keys are used to decrypt messages. The certificate authority verifies the public keys belong to whom they say they do. This creates a system where confidentiality, integrity and accountability exist for both users. In this system it is possible to also verify that tampering has not occurred by asking the sender of the message to also send a hash of the plain text message. The receiver will then rehash the sent message to verify the hashes match [Example of encryption].

During a PKI sign in [Fig. 3] a user will enter their common access card into the card reader. The reader will publish the certificate to the browser. The browser will present the certificate to the PKI server. Once the PKI sever receives the certificate an Online Certificate Status Protocol validation occurs [6]. A Certificate Revocation List [6] will be checked to verify the certificate is still good. If the certificate authority responds that the certificate is valid, the user will be allowed to access the protected applications.

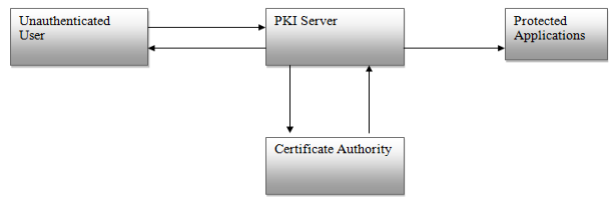

Fig. 3 – PKI Initial Sign-In

### A. Malicious Ciphers

The Prometheus Flight Simulator project will follow the NIST SP 800-61 [7] recommendations. A system will be implemented that performs heuristic virus scanning on individual severs at start up. Scans will kick off again at off peak hours and at a scheduled high risk times outlined by the security team. A network scan will take place twice daily. Audits will be conducted weekly to verify the network is running as expected. The scans will look for the most likely forms of attack first, such as Trojans, viruses and worms. Additional forms of attacks to look for can be added by the security team as needed.

Incident responses will be handled according to Department of Defense standards. Evidence will be gathered and documented about the incident. The report will then be handed off to the suitable security teams for further investigation. User education will be conducted monthly to make them aware of emerging threats and to re-enforce existing threats. Online modules will be used to ensure that the end users are receiving the training.

A bulletin system will be set up to send out emails as they occur on new issues that arise. This system will be managed by the security administrator. They will be responsible for passing along new information as it occurs. Auto run on all desktops will be disabled. Strict guidelines on the use of USB devices will be outlined and enforced. Recovery will be handled through a last known good backup. The last known good backup will be used if any files are compromised. Industry standard recovery techniques will be used if deletion of critical files that were not backed up occurs.

### B. Harden the System

Hardening will occur by using a least privileged user approach. Users will only have the minimum authorization needed to complete specific jobs. Added levels of privilege will be granted based on change management approval and a security review. Users will be monitored to ensure that they are not trying to gain higher levels of authorization then they normally have. If a user is identified in the logs trying to gain access to a protected resource above their authorization and if this incident occurs three or more times, an investigation will be kicked off. A notice of investigation will be sent to the security team.

During the hardening process unused ports will be locked down on all hardware. A port scanning tool will be used to verify the ports are closed. Regression testing will take place to verify that everything continues to work as expected. An audit will take place to verify that all sign on credentials have been changed and follow the Department of Defense guidelines on password strength.

### C. Database Design

The Oracle database [Fig. 4] will act as a repository for unsecure, secure, confidential and secret data. The data will be used by the various applications to display information to the users. While the end users must user a secure sign in to get to each application, an analysis of the application query architecture should be evaluated to ensure that other vectors of attack do not exist. If a failure of single sign on occurs, we must ensure the database is not compromised.

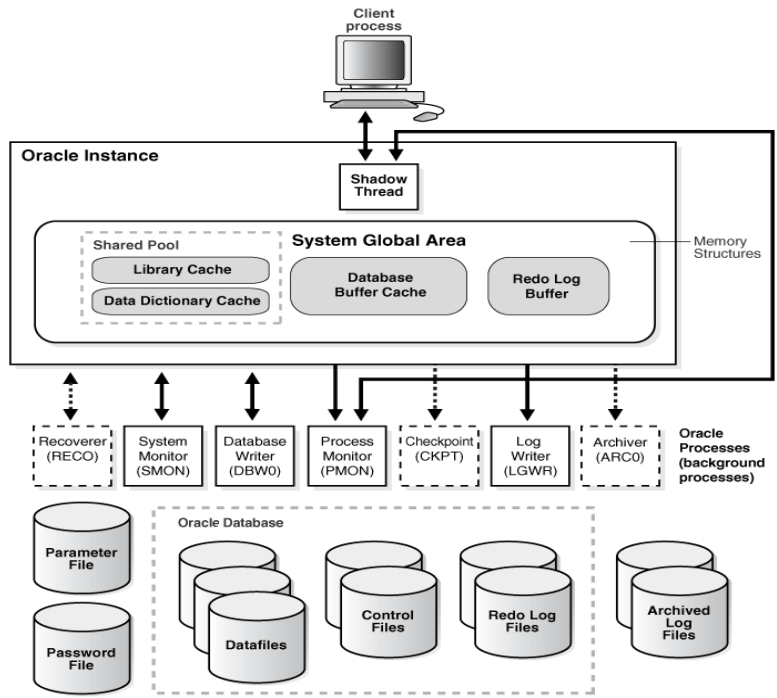

Fig. 4 Oracle Database [27]

### D. Analysis: Prometheus Flight Simulator

During development, the Oracle database will be accessed by the development team and the system administrators daily. Changes are expected to occur as needed to deploy the application for use by the end users. The database will support three applications for the baseline install. The Prometheus Flight Simulator will be the primary application querying the database. Since the applications will be internal and we have chosen a single sign on solution, there will be no need for users to enter their credentials again when they reach the applications page. This will be true to gain access to the Flight Simulator and the Back Office. The Administrator Application will use a separate security approach.

### E. Administrator Server

Administrator server functions will be treated as special functions. Access control lists will decide who can see the links to certain functions. For example the Oracle back office functions will only be accessible by users in the Oracle administrators group. Administrator user access is required to meet NIST level four requirements outlined below:

- Hard common access cards required.
- FIPS 140-2 required.
- Protection against on-line guessing.
- Protection against replay attacks.
- Protection against eavesdropping attacks.
- Protection against verifier impersonation.
- Protection against man-in-the-middle attacks.
- Protection against hijacking [11].

Since the user must meet all of these requirements before ever getting to the database software, we can be assured that we have a strong understanding of whom the user is. However, if someone can bypass all of these safeguards we will need to deal with several other forms of attacks at the application layer.

## F. Application Security

A mixture of COT's software and custom code will be developed to link one system to another. Each system will have to be assessed to verify that it meets the current FISMA [3] requirements. Application security plays a critical role in the architecture of the Prometheus Flight Simulator project. Since each application links directly with the Oracle server, steps will be taken to deal with threats such as dynamic SQL and poor coding practices that may lead to breaches at the database level. Parameters [12] will be used for all application links to avoid dynamic SQL issues. Even though the parameters will be known throughout the application, steps will be taken to ensure that user input is valid.

JavaScript will be used on the client side to verify all user input before a submission to the server. Once the input is submitted, the input will be validated on the server side. A transaction block will kicked off at this time to start the database query. If at any point the transaction fails or the user input changes, the transaction will fail. Logging will be in place to report all transaction failures. Additionally, tools will be used during the SDLC to ensure the code follows best practices and meets security requirements. CodePro Analytix will be used throughout each revision of the code base to ensure that best practices are being followed and that security errors are not occurring. CodePro Analytix will provide the following for database transaction analysis:

- Avoid building queries from user input.
- Close connections where created.
- Close order.
- Close result set where created.
- Close statement where created.
- Check that all database connections are non-static.
- Checks the source for the database connection.
- Reuse data sources for JDBC connections.
- Protect against SQL Injection.
- Help with statement creation [13].

In addition to static analysis we will also perform fuzz testing to verify user input in each of the applications. This will give us a clear picture of how the application will act with a variety of invalid inputs. The team has chosen the CERT Basic Fuzzing Framework to perform all fuzz testing on the user sign in and application interfaces. The tool will also be used on the out of the box user inputs for all commercial off the shelf products [14]. As a final precaution, a certified ethical hacker will perform a penetration test on the application. They will be tasked with getting in, gaining the highest privilege they can and then returning the full result of the user name table.

## G. Software Development Life Cycle (SDLC)

An agile [15] software development life cycle will be used throughout development. Static analysis, fuzz testing, and automated tests will be run at all iterations of the development. Code will be maintained in a Subversion repository. The code will be updated by individual developers in their branches. As components are finished, the code will be merged down to the trunk. A build master will be assigned to the code base. They will be responsible for performing weekly builds. If at any point the build fails, a bug bash will be called. The bug bash will consist of deciding the reason for the build failure. If the failure is because of a hole in the build documentation, then the development team will spend the time updating the build procedure. Otherwise, time will be spent coding a solution to the build issue.

Bugzilla will be used for end user and internal bug tracking. The team will document fixes to each bug in the tracking system. The bug number will also be documented in subversion for reference. As components are released to the trunk in subversion and after unit testing of the build has completed, the code will be tagged. The tag will act as a roll back point for each component. Bug tracking, version control, weekly builds, testing and security analysis will make up the revision control system for the Prometheus Flight Simulator project. The customer will have view only access to the code base and security analysis. Since the tools are the approved tools used by our development team, these will be the official tools for the project. The customer shall not dictate tools that will be used for software development.

## H. Software Quality, Reliability and Security

"Software quality measures how well software is designed and how well the software conforms to that design [16]." "Software reliability is the probability of failure-free operation of a computer program in a specific environment at a specific time [16]. "Security is the ability of a system to protect information and system resources with respect to confidentiality and integrity [17]. Quality and reliability are often used interchangeably. However, this is incorrect [16]. As the definitions above describe, they are different categories when it comes to software development. The categories should be considered individually during the software development lifecycle. The categories should not only encompass the software, but the entire solution. Hardware can be considered separately, but it should meet the same criteria the software is held to.

Software security should be considered its own category and be evaluated during all phases of the development. During the evaluation of quality and reliability, a security evaluation should also occur around these categories to verify they are falling into the security constraints set by the project. The Prometheus Flight Simulator Project will use the following impact attributes to determine software quality, reliably and security at a high level.

- Functionality
  - o Suitability – Was the solution used suitable for the problem? Did it the solution meet the requirements?
  - o Accuracy – Does the solution cover all the requirements outlined for this portion of the deliverable?
  - o Interoperability – Does all the hardware function together as we expect? Have all the test cases passed for the software and hardware?
- Functionality Compliance – Does the functionality meet legal regulations? For instance, are the web pages 508 compliant? [18]
- Reliability
  - o Maturity – Have the number of bugs decreased over the time we have been testing? Is the software maturing at a pace acceptable with industry standards?
  - o Fault Tolerance – How does the system handle a crash? What strain can the system take before a crash occurs?

- o Recoverability – Can the system be recovered after a crash? Can the system be restored to full operational capacity within the timelines outline within the requirements?
- o Reliability Compliance – Does the software reliability meet the compliance outlined by state and federal law?
- Usability
  - o Understandable – Is the software understandable to the end user? Do they know what the new software buys them as a work force?
  - o Learnable – Is the software easy to learn? Does the training material give enough information? Does training material exist on all areas?
  - o Operability – Do each of the user interactions do what they are reported to do? Are there multiple ways to operate the software? Do the operations work equally as well regardless of the interface? For example, if the user A chooses a graphic interface and user B chooses the command line; will the operations performed result in the same result for the user?
  - o Attractiveness – Does the look and feel of the software meet the company standards? Have industry standards been followed in the creation of web forms and other user inputs?
  - o Usability Compliance – Did the usability meet all state and federal requirements?
- Efficiency
  - o Time behaviour – Does the software start, stop, query, create, delete and display in the time required according to the requirements?
  - o Resource Utilization – Is resource use within the tolerance levels outlined for the design? Does the hardware support the software use requirements?
  - o Efficiency Compliance – Does the efficiency comply with all state and federal laws about efficiency?
- Maintainability
  - o Analysability – Does the software have logging and analysis built in according to the requirements?
  - o Changeability – Does the software have a well defined change management system with an upgrade timeline outlined?
  - o Stability – Is the software stable? Is the hardware stable? Are there activities outlined around stability maintenance?
  - o Testability – Are regression tests defined and automated? Is a test schedule outlined? Are regression tests being performed at all phases of the SDLC?
  - o Maintainability Compliance
  - [19]
- Security
  - o Confidentiality – Do only the correct personnel have access to the protected information or resources?
  - o Integrity – When the information is displayed, can we ensure that it has not been tampered with?
  - o Availability – Can the information be accessed at all times required in the requirements?
  - o Legal Compliance – Does the software security meet all the state and federal laws mandated for the project?

## I.  Encryption Mechanisms

The Prometheus Flight Simulator Project will use SSL [5] as the chosen encryption for all web connections.  The SSL connection will be created during either a PKI [5] or SecurID [20] two factor authentications.  For database transactions between internal systems, "trusted contexts and trusted connections" [21] will be used to strengthen the security when an end user queries the database or another internal system.  For each trusted context, a trusted context statement [22] will be used to define the encryption needed to establish the connection. If that encryption cannot be achieved then the connection will not take place. The end user will receive an error page and be given information to contact the administrator in this case.

The current levels of encryption available are "no encryption, low encryption, and high encryption" [22].  For all external connections, a high level of encryption or SSL 3.3 encryption will be required. The SSL 3.3 connection will use a minimum of 256-bit encryption for all external connections and a minimum of 128-bit for all secret or higher internal data transmissions [28].

## J.  Cipher Attacks and Defense

Many cipher attacks will have to be outlined and mitigated during this portion of the design. The team is specifically concerned with the following cipher attackers.

- Known Plaintext and Ciphertext-Only Attacks
- Chose Plaintext and Chosen Ciphertext Attacks
- Adaptive Chosen Plaintext and Adaptive Chosen Ciphertext Attacks
- Side Channel Attacks
- Brute Force Attacks
- Meet-in-the-Middle Attack
- Linear Cryptanalysis and Differential Cryptanalysis
- Birthday Attacks [23].

The Prometheus Flight Simulator Project will mitigate each of these threats through architectural design and by following best practices for each part of the implementation.  Steps will be taken to prevent the attacker from being able to enumerate the information needed to launch each of these attacks. For example a side channel attack would require an attacker to have a great deal of information about the hardware with the plaintext or cipher text and possibly the cryptographic algorithm [23].

   Preventing the attacker from gaining these details will be achieved by network monitoring and education. The full implementation details of the system should not be released to the public. All upgrades to the system will remain confidential. Administrators will be responsible for locking down the system so implementation details will not be accessible to anyone except the administrator. Extra steps will be taken to prevent the following attacks that may also allow the attacker to gain root access.

- Buffer overflow attacks
- Cross-site scripting
- Invalidated input
- Injection attacks
- Improper error handling [24].

The following attacks will be mitigated based on code implementation and web access design. Buffer overflow attacks will be handled by ensuring that best practices are followed during the creation of objects that may force a buffer overflow error. Cross-site scripting will be mitigated by verifying all user input originated from the user. Verifying that URL's have not been rewritten and examining all cookies to verify the correct changes are taking places as the user steps through the application.

All input will be validated to ensure that special characters have not been added to prevent a possible injection attack. Timestamps and hashes will be examined to verify that data has not been tampered with and that stale data is not being sent to the system. Error handling will be performed in a way that will give the end user the proper feedback for their transaction, but will not allow implementation details to be leaked to a would be attacker.

## K. CIA, Non-repudiation and Authenticity

The Prometheus Flight Simulator project will provide Confidentiality, Integrity and Availability through various implementations of systems and resources locally and on the network. Each user will have access during peak operation hours; integrity is maintained through a system of checks that ensure the data has not been tampered with. The checks will use hashing to verify the hash is unaltered through the transfer from client to host and host to client. Confidentiality is ensured through a minimum of RSA-1024 encryption algorithms [8]. The algorithm with a salt [9] will ensure the user data is protected in a way that makes it unlikely that a malicious hacker could gain access. Authenticity is achieved through hashing checks that determine if the encrypted data has been tampered with [28].

Non-repudiation will be achieved through asymmetric cryptography. "In some circumstances, the digital signature can be used to establish non-repudiation. If Bob can demonstrate that only Alice holds the private key, Alice cannot deny generating the signature. In general, Bob will need to rely on a third party to attest that Alice had the private key [29]".

Availability strengthened based on redundant systems in place and a strong administrative team that can react quickly to denial of service type attacks. Availability has also been heightened by using JAVA programming approaches that prevent IP's addresses from making concurrent connections over a specified time period [26]. Monitoring services will be implemented to verify where a connection is coming from. A two way trust will be established for each connection to ensure the connection will fail if trust cannot be established. If a trust for the specific user cannot be established after three tries a twenty four hour cool down period will take place for that user's IP. The administrator will receive a report of all active cool downs and their IP addresses.

IP filtering will be implemented to create a white list of IP addresses. Addresses not on the white list will never achieve access to the system. In fact the application will only perform a simple query of the current user IP against the white list. The application will immediately kick the end user out and lock out their IP for twenty four hours if they do not appear on the list. Availability has been defined as high throughout normal government work hours and low throughout nonworking hours. The administrators will perform all upgrades and maintenance during the off-hours to ensure that high availability is maintained. Backups are preserved and rolled into an available state based on any issues that may arise with the data. The data is hashed and analysed to ensure that tampering has not occurred either during storage, transmission or retrieval.

## 3. CONCLUSIONS

The government organization had a need of a sign in system that met specific criteria for their mission. Our team was tasked with creating a plan that would encompass each of their requirements. Since the team understands that this system will eventually be evaluated by the IAM and IEM [25], we have designed the system with the CIA triad in mind. Thorough analyses of PKI authentication and token authentication were conducted to give the customer an overview of how their system will be secured.

An analysis of cryptographic attack techniques were conducted to determine the threat level that this project was likely to face. Mitigation plans were documented for each portion of the threat analysis. The solution outlined takes into account each of the customer requirements. When designing this solution we sought to achieve a solution that could be leveraged moving forward for not only this organization but other organizations that were experiencing this business problem. The DoD has several specific issues they were facing such as wiretapping [33] that we had to look at separately. However, this has given us a more thorough understanding of the issues that Federal departments may face in the future. The Department of Defense has determined the solution meets all of their requirements and will be taking our recommendation around security risk management [32] for further review. We as a team feel that this solution is a win for all parties.

## Authors

**Jason Slaughter** is a graduate student in Information Security and Assurance at Capella University. He has experience as a Software Engineer, Fire Fighter, EMT, U.S. soldier and Pharmacy Technician. Jason's interests include security, software engineering and computer forensics. Jason is an avid reader and writer; he is currently working on a fantasy novel in his off time. He looks forward each day to working with all things computer related, especially in the information security sector. 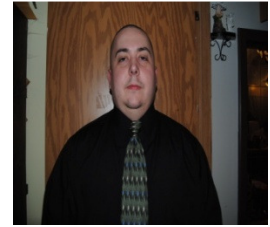

**Syed (Shawon) M. Rahman** is an assistant professor in the Department of Computer Science and Engineering at the University of Hawaii-Hilo and an adjunct faculty of information Technology, information assurance and security at the Capella University. Dr. Rahman's research interests include software engineering education, data visualization, information assurance and security, web accessibility, and software testing and quality assurance. He has published more than 50 peer-reviewed papers. He is a member of many professional organizations including ACM, ASEE, ASQ, IEEE, and UPE. 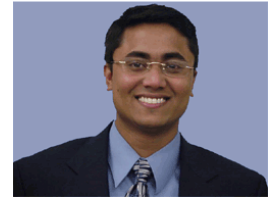